# A Hybrid cryptosystem based on Vigenère cipher and Columnar Transposition Cipher

Quist-Aphetsi Kester, MIEEE, Lecturer Faculty of Informatics, Ghana Technology University College, PMB 100 Accra North, Ghana
Phone Contact +233 209822141 Email: kquist-aphetsi@gtuc.edu.gh /  kquist@ieee.org

## Abstract

Privacy is one of the key issues addressed by information Security. Through cryptographic encryption methods, one can prevent a third party from understanding transmitted raw data over unsecured channel during signal transmission. The cryptographic methods for enhancing the security of digital contents have gained high significance in the current era. Breach of security and misuse of confidential information that has been intercepted by unauthorized parties are key problems that information security tries to solve.

This paper sets out to contribute to the general body of knowledge in the area of classical cryptography by developing a new hybrid way of encryption of plaintext.  The cryptosystem performs its encryption by encrypting the plaintext using columnar transposition cipher and further using the ciphertext to encrypt the plaintext again using Vigenère cipher. At the end, cryptanalysis was performed on the ciphertext. The implementation will be done using java programming.

## Introduction

The widen handling of digital media for information transmission through secure and unsecured channels exposes messages sent via networks to intruders or third parties. Encryption of messages in this modern age of technology becomes necessary for ensuring that data sent via communications channels become protected and made difficult for deciphering. [1]Enormous number of transfer of data and information takes place through internet, which is considered to be most efficient though it's definitely a public access medium. Therefore to counterpart this weakness, many researchers have come up with efficient algorithms to encrypt this information from plain text into ciphers [2].

In information security, encryption is the process of transforming information using an algorithm to make it unreadable to anyone except those possessing special knowledge, usually referred to as a key. The result of the process is encrypted information. The reverse process is referred to as decryption [3]. There two main algorithmic approaches to encryption, these are symmetric and asymmetric. Symmetric-key algorithms [4] are a class of algorithms for cryptography that use the same cryptographic keys for both encryption of plaintext and decryption of cipher text. The keys may be identical or there may be a simple transformation to go between the two keys. The keys, in practice, represent a shared secret between two or more parties that can be used to maintain a private information link [5]. This requirement that both parties have access to the secret key is one of the main drawbacks of symmetric key encryption, in comparison to public-key encryption. Typical examples symmetric algorithms are Advanced Encryption Standard (AES), Blowfish, Tripple Data Encryption Standard (3DES) and Serpent [6].

Asymmetric or Public key encryption on the other hand is an encryption method where a message encrypted with a recipient's public key cannot be decrypted by anyone except a possessor of the matching private key, presumably, this will be the owner of that key and the person associated with the public key used. This is used for confidentiality. [7]. Typical examples of asymmetric encryption algorithms are  Rivest Shamir Adleman (RSA),Diffie–Hellman key exchange protocol and Digital Signature Standard(DSS), which incorporates the Digital Signature Algorithm (DSA)

Modern day cryptography entails complex and advance mathematical algorithm are applied to encryption of text and cryptographic techniques for image encryption based on the RGB pixel displacement where pixel of images are shuffled to obtained a cipher image [8].

This research is aimed at contributing to the general body of knowledge in the area of the application of cryptography by employing a hybrid method of encryption. This method combines Vigenère cipher and columnar transposition cipher in its encryption process. The paper has the following structure: section II consist of related works, section III of the methodology, section IV The algorithm section V Implementation, section VI Results and Analysis and section VII concluded the paper.

## Related Works

Caesar cipher, also known as the shift cipher, is one of the simplest and most widely known classical encryption techniques. It is a type of substitution cipher in which each letter in the plaintext is replaced by a letter some fixed number of positions down the alphabet. For example, with a shift of 3, A would be replaced by D, B would become E, and so on. The encryption step performed by a Caesar cipher is often incorporated as part of more complex schemes, such as the Vigenère cipher, and still has modern application in the ROT13 system. As with all single alphabet substitution ciphers, the Caesar cipher is easily broken and in modern practice offers essentially no communication security.[9]

The encryption can also be represented using modular arithmetic by first transforming the letters into numbers, according to the scheme, A = 0, B = 1... Z = 25. [10] Encryption of a letter by a shift n can be described mathematically as, [11]

$E_n(x) = (x + n) \mod 26$

Decryption is performed similarly,

$D_n(x) = (x - n) \mod 26$

The Vigenère cipher is a method of encrypting alphabetic text by using a series of different Caesar ciphers based on the letters of a keyword. It is a simple form of polyalphabetic substitution [12][13]. The Cipher spoils the statistics of a simple Caesar cipher by using multiple Caesar ciphers. The technique is named for its inventor, Blaise de Vigenère from the court of Henry III of France in the sixteenth century, and was considered unbreakable for some 300 years [14].

Vigenère can also be viewed algebraically. If the letters A–Z are taken to be the numbers 0–25, and addition is performed modulo 26, then Vigenère encryption E using the key K can be written,[15]

$C_i = E_K(M_i) = (M_i + K_i) \mod \{26\}$

and decryption D using the key K,

$M_i = D_K(C_i) = (C_i - K_i) \mod \{26\}$,

whereas $M = M_0 \ldots M_n$ is the message, $C = C_0 \ldots C_n$ is the ciphertext and $K = K_0 \ldots K_m$ is the used key.

Thus Given m, a positive integer, $P = C = (Z_{26})^n$, and $K = (k_1, k_2 \ldots k_m)$ a key, we define:

Encryption:

$e_k(p_1, p_2 \ldots p_m) = (p_1+k_1, p_2+k_2 \ldots p_m+k_m) \pmod{26}$

Decryption:

$d_k(c_1, c_2 \ldots c_m) = (c_1-k_1, c_2-k_2 \ldots c_m-k_m) \pmod{26}$

Example:
Plaintext:    C R Y P T O G R A P H Y
Key:          L U C K L U C K L U C K
Ciphertext:   N L A Z E I  I B L J J I

A modified form of the Vigenère cipher, the alpha-qwerty cipher extended the original 26 character Vigenère cipher to a 92 characters case sensitive cipher including digits and some other symbols commonly used in the English language and can be written from a computer keyboard. The alpha-qwerty cipher also changes the mapping sequence used in the Vigenère cipher. The mapping takes from an extended alphabet sequence to extended qwerty keyboard sequence. To decrypt the code reverse mapping takes place (compliment of encryption) that is from extended QWERTY key-board to extended alphabet sequence. In short this proposed version extends and rearranges the original Vigenère table, therefore making it much more complex than the existing one. The greater character set allows more type of messages to be encrypted like passwords. It also increases the key domain and hence provides more security [16].

|   | A | B | C | D | E | F | G | H | I | J | K | L | M | N | O | P | Q | R | S | T | U | V | W | X | Y | Z |
|---|---|---|---|---|---|---|---|---|---|---|---|---|---|---|---|---|---|---|---|---|---|---|---|---|---|---|
| A | A | B | C | D | E | F | G | H | I | J | K | L | M | N | O | P | Q | R | S | T | U | V | W | X | Y | Z |
| B | B | C | D | E | F | G | H | I | J | K | L | M | N | O | P | Q | R | S | T | U | V | W | X | Y | Z | A |
| C | C | D | E | F | G | H | I | J | K | L | M | N | O | P | Q | R | S | T | U | V | W | X | Y | Z | A | B |
| D | D | E | F | G | H | I | J | K | L | M | N | O | P | Q | R | S | T | U | V | W | X | Y | Z | A | B | C |
| E | E | F | G | H | I | J | K | L | M | N | O | P | Q | R | S | T | U | V | W | X | Y | Z | A | B | C | D |
| F | F | G | H | I | J | K | L | M | N | O | P | Q | R | S | T | U | V | W | X | Y | Z | A | B | C | D | E |
| G | G | H | I | J | K | L | M | N | O | P | Q | R | S | T | U | V | W | X | Y | Z | A | B | C | D | E | F |
| H | H | I | J | K | L | M | N | O | P | Q | R | S | T | U | V | W | X | Y | Z | A | B | C | D | E | F | G |
| I | I | J | K | L | M | N | O | P | Q | R | S | T | U | V | W | X | Y | Z | A | B | C | D | E | F | G | H |
| J | J | K | L | M | N | O | P | Q | R | S | T | U | V | W | X | Y | Z | A | B | C | D | E | F | G | H | I |
| K | K | L | M | N | O | P | Q | R | S | T | U | V | W | X | Y | Z | A | B | C | D | E | F | G | H | I | J |
| L | L | M | N | O | P | Q | R | S | T | U | V | W | X | Y | Z | A | B | C | D | E | F | G | H | I | J | K |
| M | M | N | O | P | Q | R | S | T | U | V | W | X | Y | Z | A | B | C | D | E | F | G | H | I | J | K | L |
| N | N | O | P | Q | R | S | T | U | V | W | X | Y | Z | A | B | C | D | E | F | G | H | I | J | K | L | M |
| O | O | P | Q | R | S | T | U | V | W | X | Y | Z | A | B | C | D | E | F | G | H | I | J | K | L | M | N |
| P | P | Q | R | S | T | U | V | W | X | Y | Z | A | B | C | D | E | F | G | H | I | J | K | L | M | N | O |
| Q | Q | R | S | T | U | V | W | X | Y | Z | A | B | C | D | E | F | G | H | I | J | K | L | M | N | O | P |
| R | R | S | T | U | V | W | X | Y | Z | A | B | C | D | E | F | G | H | I | J | K | L | M | N | O | P | Q |
| S | S | T | U | V | W | X | Y | Z | A | B | C | D | E | F | G | H | I | J | K | L | M | N | O | P | Q | R |
| T | T | U | V | W | X | Y | Z | A | B | C | D | E | F | G | H | I | J | K | L | M | N | O | P | Q | R | S |
| U | U | V | W | X | Y | Z | A | B | C | D | E | F | G | H | I | J | K | L | M | N | O | P | Q | R | S | T |
| V | V | W | X | Y | Z | A | B | C | D | E | F | G | H | I | J | K | L | M | N | O | P | Q | R | S | T | U |
| W | W | X | Y | Z | A | B | C | D | E | F | G | H | I | J | K | L | M | N | O | P | Q | R | S | T | U | V |
| X | X | Y | Z | A | B | C | D | E | F | G | H | I | J | K | L | M | N | O | P | Q | R | S | T | U | V | W |
| Y | Y | Z | A | B | C | D | E | F | G | H | I | J | K | L | M | N | O | P | Q | R | S | T | U | V | W | X |
| Z | Z | A | B | C | D | E | F | G | H | I | J | K | L | M | N | O | P | Q | R | S | T | U | V | W | X | Y |

**Figure 1. The Vigenère square**

The algebraic description of the extended version is similar to that of the original cipher. It uses modulo 92 instead of modulo and cipher text $C_i$ is derived using a sequence different from plain text sequence $P_i$.

$C_i = E_K(P_i) = (P_i + K_i) \mod 92$

and decryption D,

$P_i = D_K(C_i) = (C_i - K_i) \mod 92$

where, $P = P_0 \ldots P_n$ is the message,

$C = C_0 \ldots C_n$ is the ciphertext and $K = K_0 \ldots K_m$ is the used key.

Friedrich Kasiski was the first to publish a successful general attack on the Vigenère cipher. Earlier attacks relied on knowledge of the plaintext, or use of a recognizable word as a key. Kasiski's method had no such dependencies. He published an account of the attack, but it's clear that there were others who were aware of it. Babbage was goaded into breaking the Vigenère cipher when John Hall Brock Thwaites submitted a "new" cipher to the Journal of the Society of the Arts. Thwaites challenged Babbage to break his cipher encoded twice, with keys of different length. Babbage succeeded in decrypting a sample, "The Vision of Sin", by Alfred Tennyson, encrypted according to the keyword "Emily", the first name of Tennyson's wife. Studies of Babbage's notes reveal that he had used the method later published by Kasiski [10] [17].

In cryptography, a transposition cipher is a process of encryption by which the positions held by units of plaintext are shifted according to a regular system or pattern, so that the ciphertext constitutes a permutation of the plaintext. That is, the order of the units is changed at the end of the shifting process. Mathematically, a bijective function is used on the characters' positions to encrypt and an inverse function to decrypt. The letters themselves are kept unchanged, which implies that the effect is only on their positions only, making their order within the message scrambled according to some well-defined scheme. Many transposition ciphers are done according to a geometric design [18][19].

In a columnar transposition, the message is written out in rows of a fixed length, and then read out again column by column, and the columns are chosen in some scrambled order. Both the width of the rows and the permutation of the columns are usually defined by a keyword [19]. Advanced forms of columnar encryption techniques are used for encryption in a matrix representation form [20].

Procedure for singe columnar transposition cipher:

1. Chose a key of a fixed length
2. Write the plain text row-by-row in rectangular form but with a fixed column which is equal to the chosen key.
3. Re-arrange the column into alphabetical column using the key as the determinant.
4. Read the message column-by-column.
5. The message read becomes the ciphertext.

Example let the key be GERMAN and the plain text be "defend the east wall of the castle"

Then we obtain the following table

<u>G E R M A N</u>
d e f e n d
t h e e a s
t w a l l o
f t h e c a
s t l e x x

Re-arranging the above we will obtain

<u>A E G M N R</u>
n e d e d f
a h t e s e
l w t l o a
c t f e a h
x t s e x l

The following ciphertext will be obtained: nalcxehwttdttfseeleedsoaxfseahl

Cryptanalysis on both Vigenère cipher and columnar transposition cipher in the past shown that, there is a level of difficulty in breaking the codes. Hence a combination of both will yield a very complex situation for the various cryptographic techniques. The weakness in Vigenère cipher is that the key is repeated throughout the encryption and that of the columnar transposition is that, the same alphabets still remains in the ciphertext and hence creating possibility for easy cryptanalysis.

This paper capitalized on the strengths and solved the weakness in the Vigenère cipher by using the strength of the columnar transposition. In my work the key is used to encrypt the plaintext using transposition cipher and then the resulting ciphertext is used as a key to encrypt the plaintext using Vigenère cipher. This makes the new method very resistive to cryptanalysis.

## Methodology

The method employs use of both Vigenère cipher and columnar transposition cipher in its encryption process. The ciphertext will first be operated on using columnar transposition cipher. A chosen key out of random will initiate the transposition process. At the end of the process, the resulting ciphertext then becomes a key for the Vigenère process. With the encryption process, a table of Vigenère cipher was created. The key is then used to operate on the message which is the plaintext to produce the final ciphertext.

This process will end up making the final ciphertext more difficult to be broken using existing cryptanalysis processes. A software program will be written to demonstrate the effectiveness of the algorithm using java programming language and cryptanalysis will be performed on the ciphertext.

## THE MATHEMATICAL ALGORITHM

Let P = plaintext and let X= a character and $X_p \in X$

where $X_p \in P$

Let K=key chosen out of random with a fixed length $X_k \in X$

where $X_k \in K$

For the columnar transposition :
Let i=1,2,3……m

And $X_{p1}$= first character of the plaintext

Let P =$X_{pi}$= $\begin{bmatrix} X_{p1}……………..X_{pm} \end{bmatrix}$

Let $Y_o$= first position of character $X_k$

Let $Y_l$= last position of character $X_k$

Let $X_{poi}$= first ith position of character $X_p$ wilth relation to $Y_o$

Let $X_{pli}$= last ith position of character $X_p$ wilth relation to $Y_l$

Ct=Columnar transposition

Position $X_{poi} \rightarrow Y_o$ and $X_{pli} \rightarrow Y_l$

Ct of P = $\begin{pmatrix} Y_o……………Y_l \\ X_{po1}………..X_{pl1} \\ X_{po2}………...X_{pl2} \\ . \qquad\qquad . \\ . \qquad\qquad . \\ . \qquad\qquad . \\ . \qquad\qquad . \\ X_{pom}………..X_{plm} \end{pmatrix}$

Let columns of Ct of p =$CtP_i$

Where i=1,2,3…….m and m= the last column

The ciphertext, $C_p$ = {$CtP_1$ + $CtP_2$ + $CtP_3$ …… $CtP_m$}

We then let $C_p$ = the key for the Vigenère cipher

For the Vigenère cipher we have the following process:

If the letters A–Z are taken to be the numbers 0–25, and addition is performed modulo 26, then Vigenère encryption E using the key K can be written,

$C_i = E_K (M_i) = (M_i + K_i) \mod \{26\}$

Thus Given m, a positive integer,

P = C = $(Z_{26})n$,

and

K = ($X_{k1}$, $X_{k2}$…$X$ km)

Encryption:

$e_k(X_{p1}, X p_2… X_{pm})$

= ($X_{p1}+X_{kn0}$, $X_{p2} + X_{kn1}…X_{pm} + X_{kmn}$) (mod 26)

$X_{pi}=x$ : [a, b]= {x ∈ I: a ≤ x ≥ b ,a=0, and b=25}

where i=0,1,2…n

At the end the ciphertext is then obtained as $C_i$.

## IMPLEMENTATION

The algorithm was implemented using java programming. The plaintext is encrypted using columnar transposition first and the operated on by the Vigenère cipher.

A key of a fixed length was then chosen at random and entered into the system. The plaintext was then transformed row-by-row in rectangular form but with a fixed column which is equal to the chosen key. The columns were then rearranged alphabetically using the key as the determinant. The final ciphertext was then obtained by adding column-by column into a row. The cipher text then becomes a key for the Vigenère process. With the encryption process, a table of Vigenère cipher was created. The key is then used to operate on the message which is the plaintext to produce a ciphertext. Below is the code for the encryption process. Where the user enters a key from the interface and also enters a message at the interface. The encryption is then done based on the predefined mathematical algorithm Witten above. The algorithm was demonstrated using java programming.

```
Scanner input = new Scanner(System.in);
System.out.println("Enter the plaintext");
    String plainText = input.nextLine();
 plainText = plainText.toUpperCase();
```

```
  System.out.print("Enter the key: ");
    String key = input.nextLine();
key = key.toUpperCase();
String message;
String encryptedMessage;
// Letters in the x-axis
int x=0;
// Letters in the y-axis
int y=0;
message = plainText;
encryptedMessage = "";
// To set the temp as [x][y]
char temp[][]=new char [key.length()][message.length()];
char msg[] = message.toCharArray();
// To populate the array
x=0;
y=0;
// To convert the message into an array of char
for (int i=0; i< msg.length;i++)
{
temp[x][y]=msg[i];
if (x==(key.length()-1))
{
x=0;
y=y+1;
} // Close if
else
{
x++;
}
} // Close for loop
// To sort the key
char t[]=new char [key.length()];
t=key.toCharArray();
Arrays.sort(t);
for (int j=0;j<y;j++)
{
for (int i=0;i<key.length();i++)
{
System.out.print(temp[i][j]);
}
System.out.println();
}
System.out.println();
// To print out row by row (i.e. y)
for (int j=0;j<y;j++){
// To compare the the sorted Key with the key
// For char in the key
for (int i=0;i<key.length();i++){
int pos=0;
// To get the position of key.charAt(i) from sorted key
for (pos=0;pos<t.length;pos++){
if (key.charAt(i)==t[pos]){
// To break the for loop once the key is found
break;
}
}
System.out.print(temp[pos][j]);
encryptedMessage+=temp[pos][j];
}
System.out.println();
}
System.out.println("output of the cipher1 "+encryptedMessage);
String Cipher1=encryptedMessage
   int tableRowSize = 26;
    int tableColumnSize = 26;
    int vignereTable[][] = new int[26][26];
     for (int rows = 0; rows < tableRowSize; rows++){
       for (int columns = 0; columns < tableColumnSize; columns++){
          vignereTable[rows][columns] = (rows + columns) % 26;
        }
      }
 String plainText = message;
String key1 = Cipher1;
String cipherText = "";
int keyIndex = 0;
    for (int ptextIndex = 0; ptextIndex < plainText.length(); ptextIndex++){
       char pChar = plainText.charAt(ptextIndex);
       int asciiVal = (int) pChar;
       if (pChar == ' '){
          cipherText += pChar;
          continue;
       }
       if (asciiVal < 65 || asciiVal > 90){
         cipherText += pChar;
         continue;
       }
       int basicPlainTextValue = ((int) pChar) - 65;
       char kChar = key.charAt(keyIndex);
       kChar+=1;
       int basicKeyValue = ((int) kChar ) - 65;
       int tableEntry = vignereTable[basicPlainTextValue][basicKeyValue];
       char cChar = (char) (tableEntry + 65);
       cipherText += cChar;
         if (keyIndex == key1.length())
         keyIndex = 0;
}
  System.out.println(" Final ciphertext is " + cipherText);
```

## Results and Analysis

The program written was used to encrypt a message and the result was analyzed by various methods of cryptanalysis.

The plain text = "IN THE FOREST THERE ARE MANY TREES WITH THE SAME HEIGHT"
The keyword was ="TRUE"

After applying the transposition cipher the result, below was obtained the padded character was the same as the key.

The Cipher1= HRTEMTSHSHHTNFSERNEIHMITIEE-HAARWTAETTOTREYETEEGT

Using the above as a key to encrypt the plaintext with the Vigenère cipher yielded the following below.

Final ciphertext= PEMLQYGYWZAMUJJIREI-UHZGMZIIZWIKDMHILTAXYIGKAX

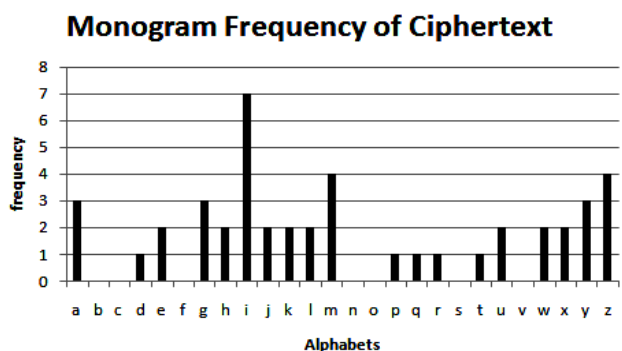

Figure 2. The frequency graph of ciphertext

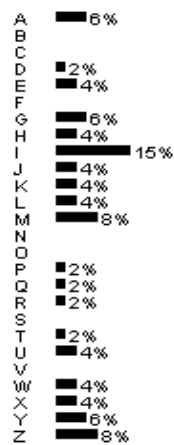

Figure 3. The Percentage of alphabet frequency within the ciphertext

The figure 2 represent the monogram frequency graph of the ciphertext and figure 3 is the percentage of each alphabet within the ciphertext.

Below is the statistical and cryptanalysis results performed on the ciphertext.

Index of Coincidence of ciphertext: The I.C. of a piece of text does not change if the text is enciphered with a substitution cipher. It is defined as:

$$I.C. = \frac{\sum_{i=A}^{i=Z} f_i(f_i - 1)}{N(N-1)}$$

Where $f_i$ is the count of letter i (where i = A, B... Z) in the ciphertext, and N is the total number of letters in the ciphertext.

**Table 1. Cryptanalysis of ciphertext**

| Cryptanalysis | Ciphertext |
| --- | --- |
| Incidence Coincidence | 0.0501 |
| Keyword Length | 1 |
| Chi-squared statistic against English distribution | 655.8480 |
| Chi-squared statistic against uniform distribution | 38.7778 |
| Statistical data: Variance | 33.9517 |
| Statistical data: Standard deviation | 5.8268 |
| Present alphabet Entropy | 3.3305 |

From the analysis the index of coincidence (IC) of the ciphertext was calculated to be 0.0501. For a normal English text of alphabet of A-Z, the Variance is normally 14.50603 and the standard deviation of 3.80868. The results indicated there is a larger deviation the ciphertext.

# Conclusion

From the analysis the index of coincidence (IC) between a string and that same string with its first few characters deleted (sometimes called a shift of the string) was obtained for A to be 0.0501 which indicates a stronger approach of the new algorithm. For a normal English text of alphabet of A-Z, the Variance is normally 14.50603 and the standard deviation of 3.80868. But from the results of the cryptanalysis, it was clearly recognized that there was a large deviation and variation from the normal observation. This indicated the strength in the new system is very high due to its hybrid nature.

# References

[1] Kester, Quist-Aphetsi. "A cryptosystem based on Vigenère cipher with varying key." International Journal of Advanced Research in Computer Engineering & Technology(IJARCET) [Online], 1.10 (2012): pp:108-113. Web. 16 Jan. 2013

[2] Kester, Quist- Aphetsi., & Danquah, Paul. (2012). A novel cryptographic key technique. In Adaptive Science & Technology (ICAST), 2012 IEEE 4th International Conference on (pp. 70-73).
[2] Abraham Sinkov, Elementary Cryptanalysis: A Mathematical Approach, Mathematical Association of America, 1966. ISBN 0-88385-622-0
[3] Nicolas Courtois, Josef Pieprzyk, "Cryptanalysis of Block Ciphers with Overdefined Systems of Equations". pp267–287, ASIACRYPT 2002
[4] Delfs, Hans & Knebl, Helmut (2007). "Symmetric-key encryption". Introduction to cryptography: principles and applications. Springer, 2007
[5] Mullen, Gary & Mummert, Carl. Finite fields and applications. American Mathematical Society. p. 112. 2007
[6] IEEE 1363: Standard Specifications for Public-Key Cryptography
[7] Kester, Q. A., & Koumadi, K. M. (2012, October). Cryptographie technique for image encryption based on the RGB pixel displacement. In Adaptive Science & Technology (ICAST), 2012 IEEE 4th International Conference on (pp. 74-77). IEEE.
[8] Bruen, Aiden A. & Forcinito, Mario A. (2011). Cryptography, Information Theory, and Error-Correction: A Handbook for the 21st Century. John Wiley & Sons. p. 21. ISBN 978-1-118-03138-4. http://books.google.com/books?id=fd2LtVgFzoMC&pg=PA21.
[9] Encryption. Wellesley college Computer Science Department lecture note retrieved from : http://cs110.wellesley.edu/lectures/L18-encryption/
[10] Caesar cipher. Retrieved from http://en.wikipedia.org/wiki/Caesar_cipher
[11] Luciano, Dennis; Gordon Prichett (January 1987). "Cryptology: From Caesar Ciphers to Public-Key Cryptosystems". The College Mathematics Journal 18 (1): 2–17. doi:10.2307/2686311. JSTOR 2686311.
[12] Bruen, Aiden A. & Forcinito, Mario A. (2011). Cryptography, Information Theory, and Error-Correction: A Handbook for the 21st Century. John Wiley & Sons. p. 21. ISBN 978-1-118-03138-4. http://books.google.com/books?id=fd2LtVgFzoMC&pg=PA21.
[13] Martin, Keith M. (2012). Everyday Cryptography. Oxford University Press. p. 142. ISBN 978-0-19-162588-6. http://books.google.com/books?id=1NHli2uzt_EC&pg=PT142.
[14] Wobst, Reinhard (2001). Cryptology Unlocked. Wiley. pp. 19. ISBN 978-0-470-06064-3.
[15] Vigenère cipher. Retrieved from http://en.wikipedia.org/wiki/Vigenère_cipher
[16] Rahmani, M. K. I., Wadhwa, N., & Malhotra, V. (2012). Advanced Computing: An International Journal (ACIJ). Alpha-Qwerty Cipher: An Extended Vigenere Cipher, 3 (3), 107-118.
[17] Franksen, O. I. (1985) Mr. Babbage's Secret: The Tale of a Cipher—and APL. Prentice Hall..
[18] Classical cipher, Transposition ciphers, Retrieved from http://en.wikipedia.org/wiki/Classical_cipher
[19] Transposition ciphers, columnar transposition Retrieved from http://en.wikipedia.org/wiki/Transposition_cipher
[20] Kester, Q.-A.; , "A public-key exchange cryptographic technique using matrix," Adaptive Science & Technology (ICAST), 2012 IEEE 4th International Conference on , vol., no., pp.78-81, 25-27 Oct. 2012

# Biographies

**QUIST-APHETSI KESTER, MIEEE:** is a global award winner 2010 (First place Winner with Gold), in Canada Toronto, of the NSBE's Consulting Design Olympiad Awards and has been recognized as a Global Consulting Design Engineer. He is a law student at the University of London UK. He is a PhD student in Computer Science. The PhD program is in collaboration between the AWBC/ Canada and the Department of Computer Science and Information Technology (DCSIT), University of Cape Coast. He had a Master of Software Engineering degree from the OUM, Malaysia and BSC in Physics from the University of Cape Coast-UCC Ghana.

He has worked in various capacities as a peer reviewer for IEEE ICAST Conference, IET-Software Journal, lecturer, Head of Digital Forensic Laboratory Department at the Ghana Technology University and Head of Computer science department. He is currently a lecturer at the Ghana Technology University College and He may be reached at kquist-aphetsi@gtuc.edu.gh or kquist@ieee.org.